\documentclass[9pt,conference]{IEEEtran}
\IEEEoverridecommandlockouts

\usepackage{cite}
\usepackage{amsmath,amssymb,amsfonts}
\usepackage{algorithmic}
\usepackage{graphicx}
\usepackage{textcomp}
\usepackage{xcolor}
\usepackage{listings} 
\usepackage{mdframed} 
\usepackage{booktabs} 
\usepackage{amssymb}
\usepackage{array}    
\usepackage{url}
\usepackage{enumitem} 

\def\tabref#1{Table \ref{#1}}
\def\figref#1{Figure \ref{#1}}
\def\secref#1{Section \ref{#1}}
\def\eqref#1{Eq (\ref{#1})}

\def\BibTeX{{\rm B\kern-.05em{\sc i\kern-.025em b}\kern-.08em
    T\kern-.1667em\lower.7ex\hbox{E}\kern-.125emX}}
\begin{document}

\title{DETECLAP: Enhancing Audio-Visual \\Representation Learning with Object Information}
\author{\IEEEauthorblockN{Shota Nakada}
\IEEEauthorblockA{
\textit{LY Corporation}\\
Tokyo, Japan \\
{\footnotesize shota.nakada@lycorp.co.jp}}
\and
\IEEEauthorblockN{Taichi Nishimura}
\IEEEauthorblockA{
\textit{LY Corporation}\\
Tokyo, Japan \\
{\footnotesize tainishi@lycorp.co.jp}}
\and
\IEEEauthorblockN{Hokuto Munakata}
\IEEEauthorblockA{
\textit{LY Corporation}\\
Tokyo, Japan \\
{\footnotesize hokuto.munakata@lycorp.co.jp}}
\and
\IEEEauthorblockN{Masayoshi Kondo}
\IEEEauthorblockA{
\textit{LY Corporation}\\
Tokyo, Japan \\
{\footnotesize masayoshi.kondo@lycorp.co.jp}}
\and
\IEEEauthorblockN{Tatsuya Komatsu}
\IEEEauthorblockA{
\textit{LY Corporation}\\
Tokyo, Japan \\
{\footnotesize komatsu.tatsuya@lycorp.co.jp}}
}

\maketitle

\begin{abstract}
Current audio-visual representation learning can capture rough object categories (e.g., ``animals'' and ``instruments''), but it lacks the ability to recognize fine-grained details, such as specific categories like ``dogs'' and ``flutes'' within animals and instruments. To address this issue, we introduce DETECLAP, a method to enhance audio-visual representation learning with object information.
Our key idea is to introduce an audio-visual label prediction loss to the existing Contrastive Audio-Visual Masked AutoEncoder to enhance its object awareness.
To avoid costly manual annotations, we prepare object labels from both audio and visual inputs using state-of-the-art language-audio models and object detectors.
We evaluate the method of audio-visual retrieval and classification using the VGGSound and AudioSet20K datasets. Our method achieves improvements in recall@10 of +1.5\% and +1.2\% for audio-to-visual and visual-to-audio retrieval, respectively, and an improvement in accuracy of +0.6\% for audio-visual classification.
\end{abstract}


\section{Introduction}
\label{sec:intro}

While visual, auditory, and linguistic modalities have traditionally been studied independently, the integration of these multiple modalities into a unified framework is gaining attention for enabling machines to have human-like integrated perception \cite{team2023gemini,lu2022unified}. 
Although it is generally costly to collect paired data among multiple modalities, paired audio-visual data can be obtained by extracting audio and visual information from the videos.
Previous work demonstrated that models trained on fused audio and visual data achieve higher performance in tasks such as retrieval and classification than those trained on single modality \cite{suris2018cross, chung2019perfect, parida2020coordinated, gong2023cav-mae, georgescu2023audiovisual}.

Building high-performance audio-visual models has become a key area of research within both the computer vision and audio processing communities. Various tasks have been proposed, such as audio-visual retrieval \cite{gong2023cav-mae} and audio-visual classification \cite{gong2023cav-mae, georgescu2023audiovisual}.
One of the important topics in audio-visual models is the learning of audio-visual representations \cite{laionclap2023, aytar2016soundnet, owens2016ambient, morgado2021audio, patrick2021compositions, wang2021multimodal} because they are effective in a wide range of downstream audio-visual tasks such as audio-visual retrieval and classification.
The Contrastive Audio-Visual Masked AutoEncoder (CAV-MAE) \cite{gong2023cav-mae} has demonstrated superior performance across multiple audio-visual tasks \cite{nagrani2021attention, jaegle2021perceiver}. The CAV-MAE is a pre-training method combining contrastive learning and the Masked Autoencoder (MAE), and it effectively learns representations from large-scale unlabeled video data. 

\begin{figure}[t]
  \centering
  \includegraphics[width=\linewidth]{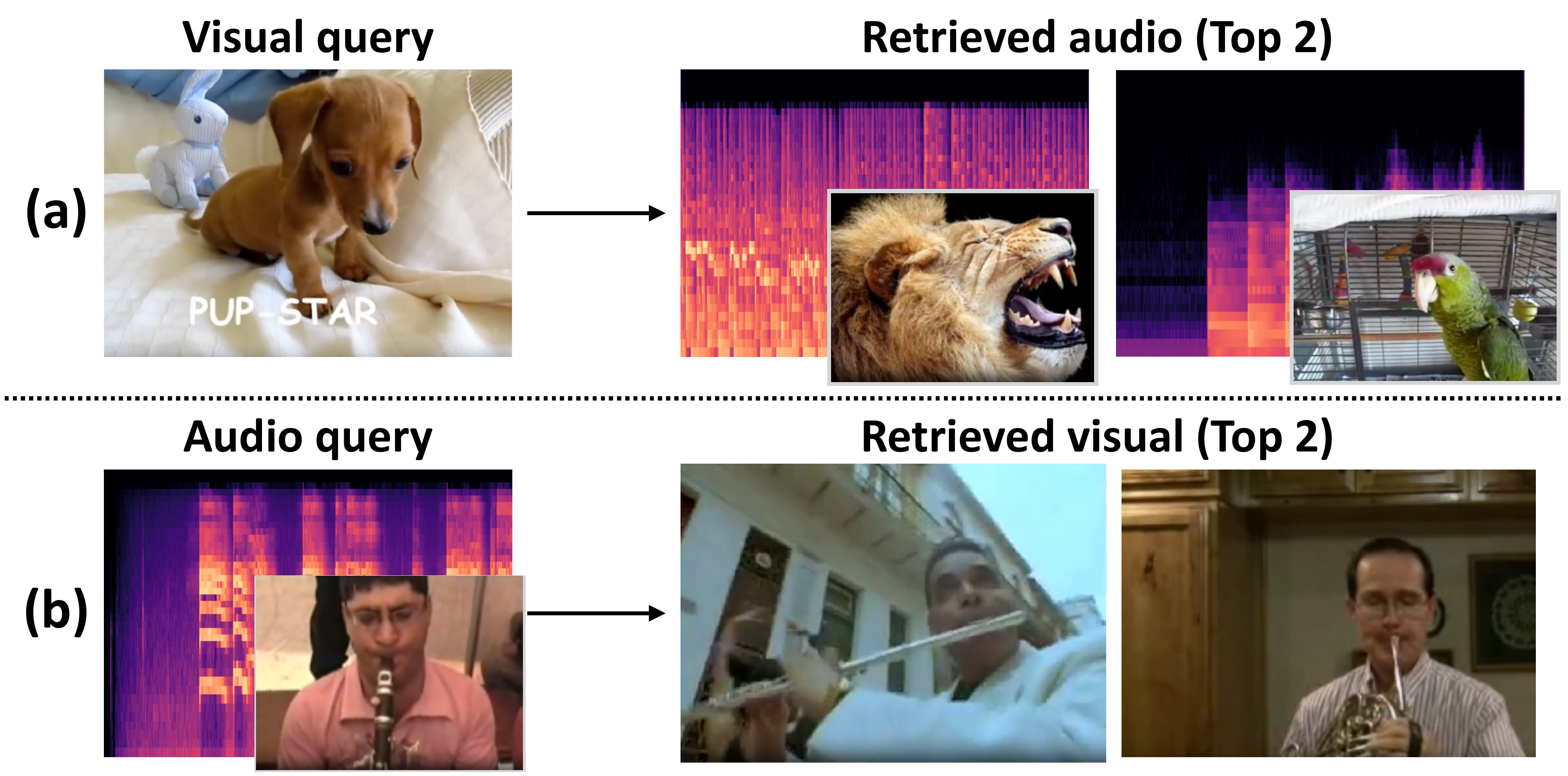}
  \caption{Failure cases of CAV-MAE in audio-visual retrieval. Although CAV-MAE captures rough category (e.g., ``animals'' and ``wind instruments''), it lacks fine-grained object information (``dogs'' and ``flute'').}
  \label{fig:cav_mae_failure_cases}
  \vspace{-4mm} 
\end{figure}

However, we observe that the CAV-MAE can capture rough object categories through pre-training based on contrastive learning and MAE, but it lacks the ability to recognize fine-grained objects, such as specific categories.
Figure \ref{fig:cav_mae_failure_cases} highlights these limitations in audio-visual retrieval.
For instance, part (a) of the figure shows that despite a visual query featuring a dog, the two most relevant audio results correspond to a lion and a bird. Similarly, part (b) shows that for an oboe audio query, the most relevant visuals depict a flute and a horn.
This limitation arises because the pre-training based on contrastive learning and the MAE only acquires representations based on the audio spectrograms and visual shapes or textures, without considering the detailed object categories in videos.

\begin{figure*}[t]
  \centering
  \includegraphics[width=0.95\textwidth]{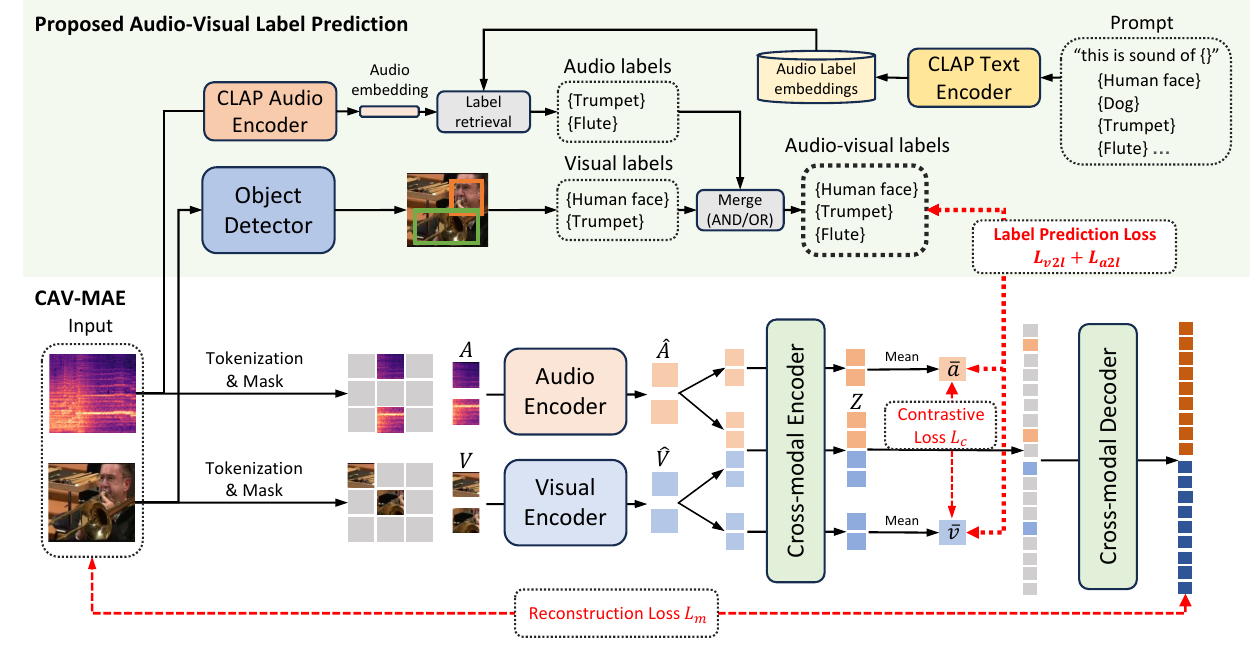}
  \caption{Overview of the proposed method DETECLAP. To enhance CAV-MAE with object information, we apply CLAP and object detector to the videos in the dataset, thereby acquiring audio-visual labels. Based on these labels, we train CAV-MAE with audio-visual label prediction loss.}
  \label{fig:cav-mae}
  \vspace{-4mm} 
\end{figure*}

%


To address this issue, we propose DETECLAP, a method to enhance audio-visual representation learning by incorporating object information. 
Our approach mitigates the issue by integrating object information from audio and visual modalities obtained from external models into CAV-MAE.
Specifically, DETECLAP introduces an audio-visual prediction loss during training, allowing the model to predict hundreds of audio-visual object labels from the extracted features.
Because manual annotation for labeling objects is costly, we automatically generate these labels using external language-audio models and object detectors. For this, we leverage state-of-the-art tools, utilizing CLAP \cite{CLAP2022} to extract audio objects and YOLOv8\cite{YOLOv8} to detect visual objects.

We evaluate our proposed method on audio-visual retrieval and audio-visual classification tasks using VGGSound \cite{chen2020vggsound} and AudioSet20K \cite{audioset}. Our method achieves improvements in recall@10 of +1.5\% and +1.2\% for audio-to-visual and visual-to-audio retrieval, respectively, and an improvement in accuracy of +0.6\% for audio-visual classification in the VGGSound dataset.

\section{Preliminary: CAV-MAE}
\label{subsec:cav_mae}

We briefly review the existing audio-visual representation learning method CAV-MAE (\figref{fig:cav-mae} bottom).
\\\textbf{(1) Tokenization and Mask.}
In CAV-MAE, audio and visual data in the input video are preprocessed separately. 
For the audio component, the input audio is first converted into $1024$ (time) $\times$ $128$ (frequency) log mel filterbank features by sliding a 25ms Hanning window every 10ms. It is then segmented into 512 $16\times16$ patches, and 75\% of the patches are randomly masked. The audio patches are represented as $\mathbf{A} = (\mathbf{a}_1, \ldots, \mathbf{a}_t, \ldots, \mathbf{a}_T) \in \mathbb{R}^{T \times 512}$, where $T$ is the number of non-masked patches.
For the visual component, a single frame is sampled from each video during training and inference. These frames are resized to $224 \times 224$, segmented into 196 patches, and 75\% of the patches are randomly masked, similar to a masked autoencoder \cite{he2022masked}. Finally, visual patches represented as $\mathbf{V} = (\mathbf{v}_1, \ldots, \mathbf{v}_k, \ldots, \mathbf{v}_{K}) \in \mathbb{R}^{K \times 256}$, where $K$ is the number of non-masked patches.
\\\textbf{(2) Audio-encoder and Visual encoder.}
Given visual and audio patches $(\mathbf{V},\mathbf{A})$, the uni-modal audio and visual encoders separately convert $\hat{\mathbf{V}}$ and $\hat{\mathbf{A}}$ into contextualized vectors $(\hat{\mathbf{V}}, \hat{\mathbf{A}})$ respectively:
\begin{align}
    \hat{\mathbf{A}} = \mathrm{AudioEncoder}(\mathbf{A}), \ \ \hat{\mathbf{V}} &= \mathrm{VisualEncoder}(\mathbf{V}),
\end{align}
where the encoders are eleven-layer Transformers \cite{vaswani2017attention}.
\\\textbf{(3) Cross-modal encoder.}
The cross-modal encoder (single-layer Transformers) concatenates $\hat{\mathbf{A}}$ and $\hat{\mathbf{V}}$ and encodes their joint representations $\mathbf{Z}$:
\begin{align}
    \label{eq:z}
    \mathbf{Z} &= \mathrm{CrossModalEncoder}([\hat{\mathbf{A}}; \hat{\mathbf{V}}]).
\end{align}
Additionally, $\hat{\mathbf{A}}$ and $\hat{\mathbf{V}}$ are separately input into the cross-modal encoder and mean-pooled to obtain the $(\bar{\mathbf{a}}, \bar{\mathbf{v}})$, which are used for audio-visual contrastive loss in \eqref{eq:contrastive} and audio-visual prediction loss in \eqref{eq:prediction_loss}:
\begin{equation}
\begin{aligned}
    \label{eq:meanpool}
    \bar{\mathbf{a}} &= \mathrm{MeanPool}(\mathrm{CrossModalEncoder}(\hat{\mathbf{A}})),\\
    \bar{\mathbf{v}} &= \mathrm{MeanPool}(\mathrm{CrossModalEncoder}(\hat{\mathbf{V}})),
\end{aligned}
\end{equation}
\textbf{(4) Cross-modal Decoder.} 
Given audio-visual joint representation $\mathbf{Z}$ obtained in \eqref{eq:z}, the cross-modal decoder reconstructs the input patches. This decoder is composed of eight-layer Transformers and the reconstructed patches are denoted as $\mathbf{A}_{r}$ and $\mathbf{V}_{r}$. This reconstruction occurs after the padded patches $\mathbf{Z}_{\text{pad}}$ are inserted. It can be represented as follows:
\begin{align}
[\mathbf{A}_{r}; \mathbf{V}_{r}] &= \mathrm{CrossModalDecoder}(\mathbf{Z}_{\text{pad}}).
\end{align}

The reconstructed features $\hat{\mathbf{Z}}$ are then used for training the masked autoencoder as shown in \eqref{eq:masked}.
\\\textbf{(5) Loss calculation.}
The overall loss $\mathcal{L}_{base}$ is computed as follows:
\begin{align}
    \mathcal{L}_c &= \mathrm{ContrastiveLoss}(\bar{\mathbf{a}}^{+}, \bar{\mathbf{v}}^{+}, \bar{\mathbf{a}}^{-}, \bar{\mathbf{v}}^{-}), 
    \label{eq:contrastive} \\
    \mathcal{L}_r &= \mathrm{MSELoss}([\mathbf{A}_{r}; \mathbf{V}_{r}],[\mathbf{A}_{orig}; \mathbf{V}_{orig}]), \label{eq:masked} \\
    \mathcal{L}_{base} &= \mathcal{L}_{c} + \mathcal{L}_{r},
\end{align}
where $\bar{\mathbf{a}}^{+}, \bar{\mathbf{v}}^{+}, \bar{\mathbf{a}}^{-}, \bar{\mathbf{v}}^{-}$ represent the positive/negative video and audio vectors sampled in the mini-batch, and $\mathbf{A}_{orig}, \mathbf{V}_{orig}$ indicate all the input patch vectors before masking.
$\mathcal{L}_c$ represents the contrastive loss \cite{chen2020simple} and $\mathcal{L}_r$ represents the reconstruction loss.

\section{method}
As shown in \figref{fig:cav-mae}, DETECLAP incorporates an audio-visual label prediction loss into CAV-MAE to enhance object awareness.
This loss is calculated using the audio-visual label $\mathbf{y}_{av}$, a fusion of the audio and visual labels from the input audio and video. In this section, we discuss training methods of the audio-visual label prediction loss in \secref{sec:label-prediction}, the methods for creating labels for each modality in \secref{sec:audio_label} and \secref{sec:visual_label}, and the merging strategies in \secref{sec:merge_label}.

\label{sec:method}

\subsection{Audio-visual label prediction loss}
\label{sec:label-prediction}

The audio-visual label prediction loss is calculated using the audio-visual labels $\mathbf{y}_{av}$. Given the mean-pooled vectors $(\bar{\mathbf{a}}, \bar{\mathbf{v}})$ in \eqref{eq:meanpool}, we add a single linear perceptron layer with weight matrices $\mathbf{W}_a, \mathbf{W}_v \in \mathbb{R}^{N \times D}$ and a sigmoid activation function $\sigma(\cdot)$ to enable the vectors $(\bar{\mathbf{a}}, \bar{\mathbf{v}})$ to recognize objects:

\begin{align}
    \label{eq:prediction_loss}
    \hat{\mathbf{y}}_a &= \sigma(\mathbf{W}_a(\bar{\mathbf{a}})),\ \
    \hat{\mathbf{y}}_v = \sigma(\mathbf{W}_v(\bar{\mathbf{v}})),\\ 
    \mathcal{L}_{a2l} &= \mathrm{BinaryCrossEntropyLoss}(\hat{\mathbf{y}}_a,\mathbf{y}_{av}), \\
    \mathcal{L}_{v2l} &= \mathrm{BinaryCrossEntropyLoss}(\hat{\mathbf{y}}_v,\mathbf{y}_{av}), 
\end{align}
where $\mathcal{L}_{a2l}$ and $\mathcal{L}_{v2l}$ are visual-to-label and audio-to-label binary cross entropy losses. We train the CAV-MAE with object information by optimizing the loss:
\begin{align}
\mathcal{L}_{deteclap}=\mathcal{L}_{base} + \mathcal{L}_{a2l} + \mathcal{L}_{v2l}.
\end{align}

\subsection{Acquiring audio labels using CLAP}
\label{sec:audio_label}
Audio labels are multi-labels representing objects contained in the audio, where each label is a hard label that takes a value of 0 or 1 for each element (object).
We obtain audio labels by following the approach used in zero-shot audio tagging with CLAP\cite{CLAP2022}, a state-of-the-art language-audio model.
The candidates for audio label names are listed in advance, and label names are input into CLAP's text encoder using the prompt "this is sound of \{label name\}" to obtain text embeddings. The input audio is then fed into CLAP's audio encoder to obtain audio embedding. The cosine similarities between the text embeddings and the audio embedding are calculated. Only labels whose similarity exceeds a threshold are adopted as the audio-label $\mathbf{y}_a \in \{0, 1\}^C$. Here, the threshold is a predefined constant set between $[0, 1]$, $C$ represents the number of label categories, and 1 is assigned if a certain label is detected, while 0 is assigned if it is not.

\subsection{Acquiring visual labels using object detectors}
\label{sec:visual_label}
Visual labels are multi-labels representing objects contained in the video, where each label is a hard label similar to audio labels.
They are obtained by using an object detector. In this method, we employ a state-of-the-art object detector, YOLOv8. We sample 10 frames from the video, input them into YOLOv8, and then obtain the detected objects and their probabilities. Only the labels of detected objects with probabilities exceeding the threshold are adopted as visual labels $\mathbf{y}_v \in \{0, 1\}^C$.

\subsection{Merging strategies}
\label{sec:merge_label}
While it is possible to train models independently using audio labels and visual labels, we hypothesized that sharing information from both modalities could enhance performance.
Therefore, we employ audio-visual labels, which are created by merging audio and visual labels, for training our model.
We define two types of merge operations, AND and OR, for creating audio-visual labels. The AND operation is defined as the element-wise logical AND operation on the audio labels and visual labels, while the OR operation is defined as the element-wise logical OR operation. 
The audio-visual label prediction loss is calculated using the audio-visual labels $\mathbf{y}_{av} \in \{0, 1\}^C$ obtained through the merge operation. 
We denote the models trained using labels created with AND and OR operations as \textbf{DETECLAP (AND)} and \textbf{DETECLAP (OR)}, respectively.
In addition, we compare the performance of the following models: \textbf{CAV-MAE}, original CAV-MAE; \textbf{DETECLAP (visual)}, using only visual labels; \textbf{DETECLAP (audio)}, using only audio labels; \textbf{DETECLAP (separate)}, visual features predict visual labels and audio features predict audio labels.

\section{Experiments}
\label{sec:experiments}

\begin{table}[t]
\centering
\caption{Audio-to-visual and visual-to-audio retrieval performance comparison on the VGGSound and AudioSet20K datasets.}
\setlength{\tabcolsep}{2pt} 
\scalebox{1.2}{
\begin{tabular}{@{}lcccccc@{}}

\toprule
\textbf{Method} & \multicolumn{3}{c}{\textbf{VGGSound}} & \multicolumn{3}{c}{\textbf{AudioSet20K}} \\
& \textbf{R1} & \textbf{R5} & \textbf{R10} & \textbf{R1} & \textbf{R5} & \textbf{R10} \\
\midrule
\multicolumn{7}{l}{\textit{\textbf{audio-to-visual}}} \\
CAV-MAE & 15.1 & 36.6 & 48.0 & 7.4 & 18.4 & 25.7 \\
DETECLAP (visual) & 15.0 & 36.8 & 46.7 & 6.9 & 17.2 & 25.3 \\
DETECLAP (audio) & 14.5 & 38.8 & 48.6 & 6.8 & 19.4 & 27.0 \\
DETECLAP (separate) & \textbf{15.9} & 36.9 & 47.9 & 7.5 & 18.1 & 26.3 \\
DETECLAP (AND) & 15.1 & 36.1 & 47.6 & 7.2 & 18.5 & 25.9 \\
DETECLAP (OR) & 15.2 & \textbf{39.2} & \textbf{49.5} & \textbf{7.9} & \textbf{20.4} & \textbf{27.7} \\
\midrule
\multicolumn{7}{l}{\textit{\textbf{visual-to-audio}}} \\
CAV-MAE & 15.7 & 39.6 & 50.5 & 7.4 & 20.0 & 28.9 \\
DETECLAP (visual) & 16.3 & 40.6 & 51.2 & 7.4 & 20.3 & 28.1 \\
DETECLAP (audio) & \textbf{17.8} & 41.6 & \textbf{52.5} & \textbf{8.2} & 20.7 & 28.6 \\
DETECLAP (separate) & 16.7 & 40.5 & 51.8 & 7.1 & 20.3 & 28.1 \\
DETECLAP (AND) & 16.7 & 40.5 & 51.6 & 7.5 & 20.2 & 28.8 \\
DETECLAP (OR) & 17.4 & \textbf{43.0} & 51.7 & \textbf{8.2} & \textbf{22.1} & \textbf{29.7} \\
\bottomrule
\end{tabular}
}
\label{tab:retrieval}
\end{table}

\begin{table}[t]
\centering
\caption{Audio-visual classification performance comparison on the VGGSound and AudioSet20K datasets.}
\scalebox{1.2}{
\begin{tabular}{@{}lcc@{}}

\toprule
\textbf{Method} & \multicolumn{1}{c}{\textbf{VGGSound}} & \multicolumn{1}{c}{\textbf{AudioSet20K}} \\
& \textbf{Accuracy} & \textbf{mAP} \\
\midrule
ResNet50(audio) \cite{chen2020vggsound}& 51.0 & - \\
AudioSlowFast \cite{kazakos2021slow}& 52.5 & - \\
MaskSpec \cite{maskspec}& -  & 32.3 \\
Audio-MAE \cite{audiomae}& - & 37.0 \\
CAV-MAE & 58.9 & 38.4 \\
DETECLAP (visual) & 59.1 & 39.7 \\
DETECLAP (audio) & 59.3 & 38.7 \\
DETECLAP (separate) & 58.8 & 38.6 \\
DETECLAP (AND) & 58.6 & \textbf{40.0} \\
DETECLAP (OR) & \textbf{59.5} & 39.6 \\
\bottomrule
\end{tabular}
}
\label{tab:classification}
\end{table}

\begin{table}[t]
\centering
\caption{Comparison of performance for different label types in audio-to-visual and visual-to-audio retrieval tasks.}
\setlength{\tabcolsep}{2pt} 
\scalebox{1.2}{
\begin{tabular}{@{}lcccccc@{}}
\toprule
\textbf{Method} & \multicolumn{3}{c}{\textbf{audio-to-visual}} & \multicolumn{3}{c}{\textbf{visual-to-audio}} \\
& \textbf{R1} & \textbf{R5} & \textbf{R10} & \textbf{R1} & \textbf{R5} & \textbf{R10} \\
\midrule
hard label (OR) & 15.2 & \textbf{39.2} & \textbf{49.5} & \textbf{17.4} & \textbf{43.0} & 51.2 \\
soft label (audio) & 15.2 & 36.4 & 46.9 & 17.2 & 39.9 & 49.9 \\
soft label (visual) & 15.0 & 37.9 & 47.9 & 16.4 & 40.9 & \textbf{52.2} \\
soft label (max) & \textbf{15.5} & 37.7 & 47.1 & 16.8 & 40.3 & 50.1 \\
\bottomrule
\end{tabular}
}
\label{tab:performance_comparison}
\end{table}




\subsection{Experimental settings}
\textbf{Dataset.} 
We conduct evaluations using the VGGSound\cite{chen2020vggsound} and AudioSet20K datasets. VGGSound is a large-scale video dataset consisting of 200,000 videos which is annotated with 309 category labels. AudioSet20K, a subset of AudioSet\cite{audioset} which is annotated with 517 category labels. 
Since AudioSet includes unavailable videos on YouTube, we have not obtained a complete dataset.
Our models are pretrained with the training data from VGGSound.


\noindent \textbf{Tasks and evaluation metrics.} To verify the effectiveness of the proposed methods, we tackle audio-visual retrieval and audio-visual classification. We evaluate our method using the VGGSound and AudioSet20K datasets.
For audio-visual retrieval, as with previous work, we measure the performance using recall@$K$ ($K=1,5,10$), which computes the percentage of audio/video queries that correctly retrieve positive videos/audio in the top $K$ of the ranking list.
For audio-visual classification, VGGSound is utilized for both pre-training and fine-tuning, while AudioSet20K is exclusively used for fine-tuning. The manually-annotated category labels are not employed during pre-training but are utilized in the fine-tuning step. As with previous work, we measure the performance using accuracy for VGGSound and mean average precision (mAP) for AudioSet20K.

\noindent \textbf{Implementation details.} We set the model's hyperparameters to match those of CAV-MAE, such as batch size, hidden size, and epochs. We use Adam optimizer \cite{adam} to train the model with learning rate $lr=5.0\times10^{-5}$. We used the 2023 version of Microsoft CLAP for audio label extraction and YOLOv8x-oiv7 for object detection. We use the Open Images Dataset V7 which consists of 600 object classes for audio and visual object list.


\begin{figure}[t]
  \centering
  \includegraphics[width=1.00\linewidth]{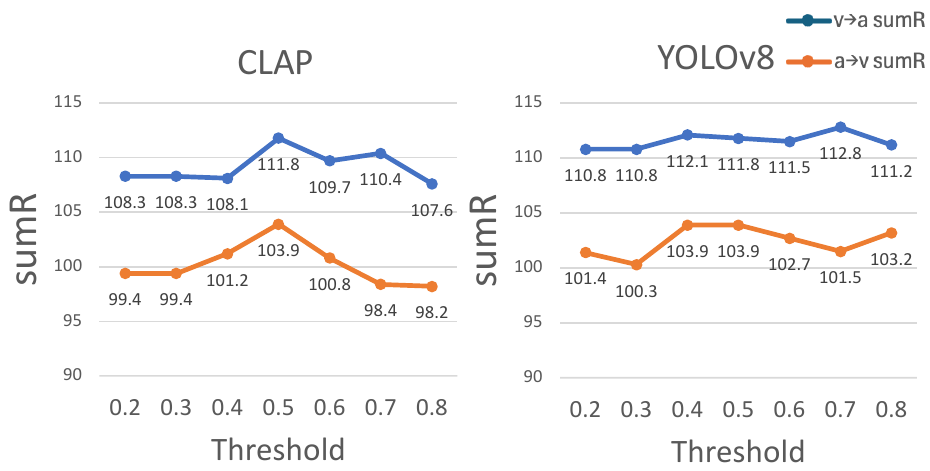}
  \caption{The sensitivity of varying thresholds on performance during audio-visual label generation. The left figure shows the performance changes when adjusting the threshold for CLAP, while the right figure details the changes for YOLOv8.}
  \label{fig:threshold}
  \vspace{-4mm} 
\end{figure}

\subsection{Results on audio-visual retrieval}
\label{sec:retrieval}
\tabref{tab:retrieval} demonstrates the performance of audio-to-visual and visual-to-audio retrieval. When comparing the CAV-MAE with DETECLAP, the latter consistently outperforms the former across all metrics. 
Specifically, it has been observed that DETECLAP (audio) performs better than DETECLAP (visual), which may be attributed to the high importance of audio information in datasets such as VGGSound and AudioSet. 
Additionally, DETECLAP (OR) shows superior performance compared to DETECLAP (AND) and DETECLAP (separate). The OR operation facilitates the transfer of object information, which can only be obtained from one modality, to the other modality, potentially enabling accurate correspondence between audio and visual modalities. These findings suggest that the choice of labels for model pre-training is crucial for enhancing performance.

\begin{figure}[t]
  \centering
  \includegraphics[width=1.05\linewidth]{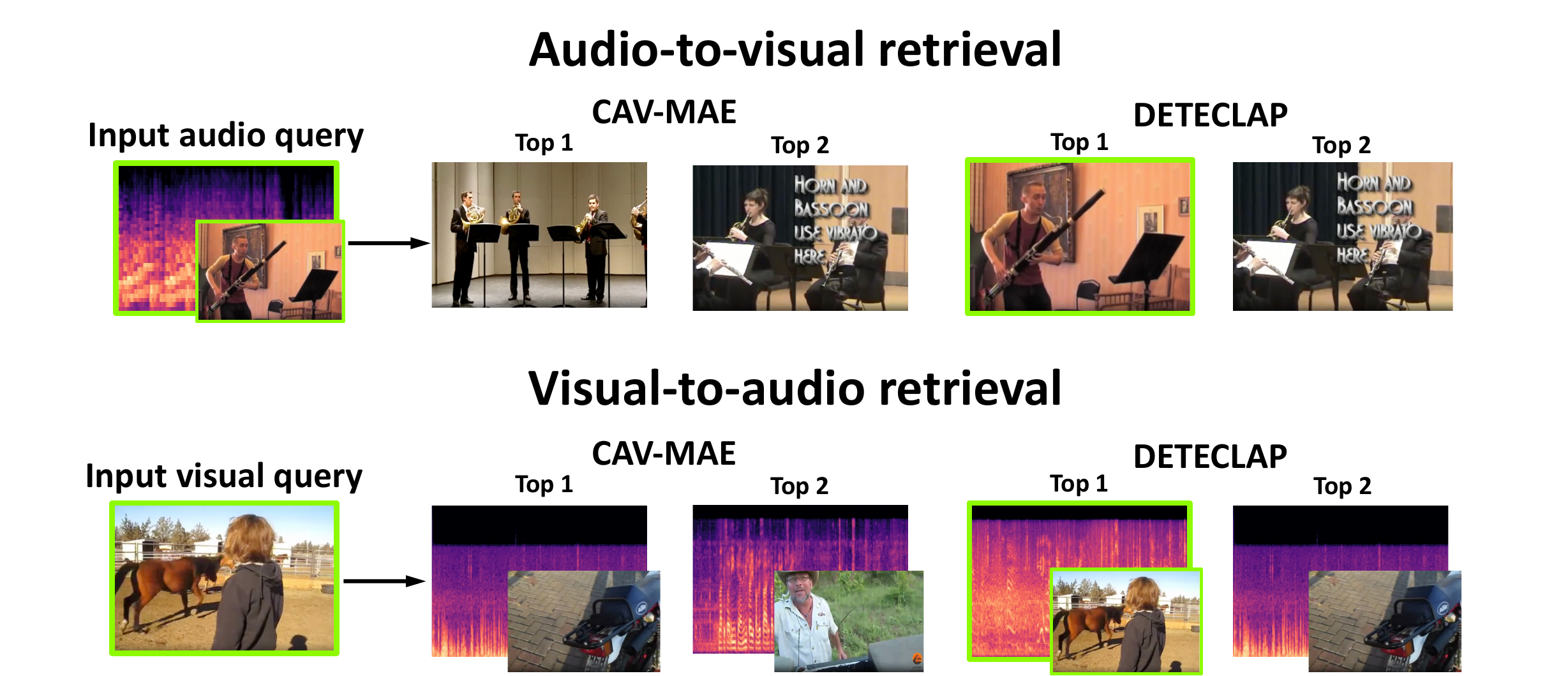}
  \caption{Retrieved visual/audio from input audio/visual queries.}
  \label{fig:retrieved_examples}
  \vspace{-4mm} 
  \label{fig:retrieved_examples}
\end{figure}

\textbf{Qualitative evaluation.}
\figref{fig:retrieved_examples} shows retrieved results of CAV-MAE and DETECLAP. In audio-to-visual retrieval, given audio of a man playing the oboe, CAV-MAE fails to retrieve the corresponding video, but our method successfully retrieves it featuring an oboe being played. In visual-to-audio retrieval, given a visual query of a horse running, CAV-MAE cannot retrieve the corresponding audio, whereas the proposed method succeeds. These results demonstrate that the proposed method enhances object recognition capabilities between audio and visual modalities, leading to improved retrieval performance.

\subsection{Results on audio-visual classification}
Table~\ref{tab:classification} shows the performance of audio-visual classification. Similar to the results in audio-visual retrieval, DETECLAP consistently outperforms CAV-MAE across all evaluated metrics. When compared with existing audio-based methods such as AudioSlowFast and AudioMAE, the proposed method significantly enhances performance. These results demonstrate that DETECLAP can effectively acquire high-quality representations for downstream tasks, emphasizing the importance of integrating audio and visual information in addressing audio-visual classification challenges.

\subsection{Comparative Study}
\noindent  \textbf{Sensitivity of threshold.}
To investigate the sensitivity of the threshold in CLAP and YOLOv8, we conduct an experiment to observe changes in sumR (the sum of Recall@{1, 5, 10}) when changing the thresholds for CLAP and YOLOv8. For this experiment, we focus on retrieval tasks using the VGGSound dataset.
During the experiment, while the threshold for one model was adjusted, the threshold for the other model was held constant at 0.5. A lower threshold indicates that labels are more likely to be adopted, whereas a higher threshold makes them less likely. According to \figref{fig:threshold}, CLAP exhibits high performance at a threshold of 0.5, while YOLOv8 shows high performance at a threshold of 0.4. These results underscore the importance of selecting appropriate thresholds for optimal model performance.

\noindent  \textbf{Soft label vs hard label.}
In our proposed method, we employed hard labels for the audio-visual label loss in \secref{sec:label-prediction}, however soft labels are also considered.
It is commonly accepted that using soft labels can more effectively distill knowledge from a teacher model \cite{hinton2015distilling}. However, some studies argue that under certain conditions, using hard labels can yield better performance\cite{hwang2023comparison}.
Therefore, we conduct an experiment to determine whether hard labels or soft labels are more appropriate in our proposed method. Table 3 compares the performance when hard labels and soft labels are applied in the retrieval task on the VGGSound dataset. We implemented three variations of soft labels: soft label (audio) using the cosine similarity of CLAP, soft label (visual) using the object probability of YOLOv8, and soft label (max) using the maximum of these two. The results indicate that in our proposed method, hard labels outperform soft labels. 
Using soft labels to align output distributions in knowledge distillation may be excessive for the purpose of injecting object information into CAV-MAE-based representation learning, potentially hindering existing representation learning processes.

\section{Conclusion}
\label{sec:conclusion}

In this paper, we introduced DETECLAP, an approach that enhances audio-visual representation learning by incorporating object information into the CAV-MAE. By integrating audio and visual object labels through the use of CLAP and YOLOv8, DETECLAP significantly improves performance on audio-visual retrieval and classification tasks.
Our experiments on the VGGSound and AudioSet20K datasets show that DETECLAP outperforms the baseline CAV-MAE when using combined audio-visual labels. 
DETECLAP advances multimodal learning by enhancing the interplay between auditory and visual modalities through object-aware learning.

\bibliographystyle{IEEEtran}
\bibliography{main}

\end{document}